\documentclass[10pt,letterpaper]{article}
\usepackage[top=0.85in,left=0.85in,right=0.85in,footskip=0.75in]{geometry}

\usepackage{amsmath,amssymb}

\usepackage{changepage}

\usepackage{textcomp,marvosym}

\usepackage{cite}

\usepackage{nameref,hyperref}

\usepackage[nopatch=eqnum]{microtype}
\DisableLigatures[f]{encoding = *, family = * }

\usepackage[table]{xcolor}

\usepackage{array}

\newcolumntype{+}{!{\vrule width 2pt}}

\newlength\savedwidth

\newcommand{\Kext}{K_{\mathrm{ext}}}
\newcommand{\thetaext}{\theta_{\mathrm{ext}}}

\raggedright
\usepackage[aboveskip=1pt,labelfont=bf,labelsep=period,justification=raggedright,singlelinecheck=off]{caption}
\renewcommand{\figurename}{Fig}

\makeatletter
\renewcommand{\@biblabel}[1]{\quad#1.}
\makeatother

\usepackage{lastpage,fancyhdr,graphicx}
\usepackage{epstopdf}
\fancyheadoffset[L]{2.25in}
\fancyfootoffset[L]{2.25in}
\begin{document}
\vspace*{0.2in}

\begin{flushleft}
{\Large
\textbf\newline{A universal animal communication tempo resonates with the receiver's brain} 
}
\newline
\\
Guy Amichay\textsuperscript{1,2,3*},
Vijay Balasubramanian\textsuperscript{4,5,6},
Daniel M. Abrams\textsuperscript{1,2,3,7*},


\bigskip

\textbf{1} Department of Engineering Sciences and Applied Mathematics, Northwestern University, Evanston, IL, USA
\\
\textbf{2} Northwestern Institute on Complex Systems, Northwestern University, Evanston, IL, USA
\\
\textbf{3} National Institute for Theory and Mathematics in Biology, Northwestern University, Evanston, IL, USA
\\
\textbf{4} David Rittenhouse Laboratory, University of Pennsylvania, Philadelphia, PA, USA
\\
\textbf{5} Santa Fe Institute, Santa Fe, NM, USA
\\
\textbf{6} Rudolf Peierls Centre for Theoretical Physics, University of Oxford, Oxford, UK
\\
\textbf{7} Department of Physics and Astronomy, Northwestern University, Evanston, IL, USA

\bigskip

%
%

*guy.amichay@northwestern.edu \\
*dmabrams@northwestern.edu

\end{flushleft}
\section*{Abstract}
During fieldwork in Thailand we observed nearly identical frequencies of co-located flashing fireflies and chirping crickets. Motivated by this, we perform a meta-analysis and show an abundance of evolutionarily distinct species that communicate isochronously at $\sim$0.5-4 Hz, suggesting that this might be a frequency ``hotspot.'' We hypothesize that this timescale may have a universal basis in the biophysics of the receiver's neurons. We test this by demonstrating that small receiver circuits constructed from elements representing typical neurons will be most responsive in the observed frequency range.


\section*{Introduction}
In the summer of 2022 members of our research group went on a field excursion to Thailand to film the \textit{Pteropyx malaccae} firefly, famous for its synchronous displays. While filming, we noticed that nearby crickets\footnote{Most likely of the \emph{Podoscirtinae} subfamily; unfortunately we failed to catch any specimen to confirm the exact species.} often seemed to chirp in synchrony with the fireflies. Luckily our microphones were on. Were the rhythmic communications at near-identical tempos (see Fig.~\ref{fig:combined}A, B) a case of interspecies, intermodal synchrony? Detailed analysis showed that the crickets and fireflies were not actually synchronized. But their signaling frequencies, both close to 2.4 Hz, differed by only about 10\%. 

If there is no synchrony, why would these two distinct species operate at such a similar tempo when they could in principle choose from a wide range of options? Note that we refer here to the \textit{temporal} properties of the signal (i.e., how often an animal calls over time) as opposed to the \textit{spectral} properties of the signal (i.e., the specific pitches in an audio signal).

\section*{Results}
\subsection*{Data}
\begin{figure*}[t!]
\centering
\includegraphics[width=\linewidth]{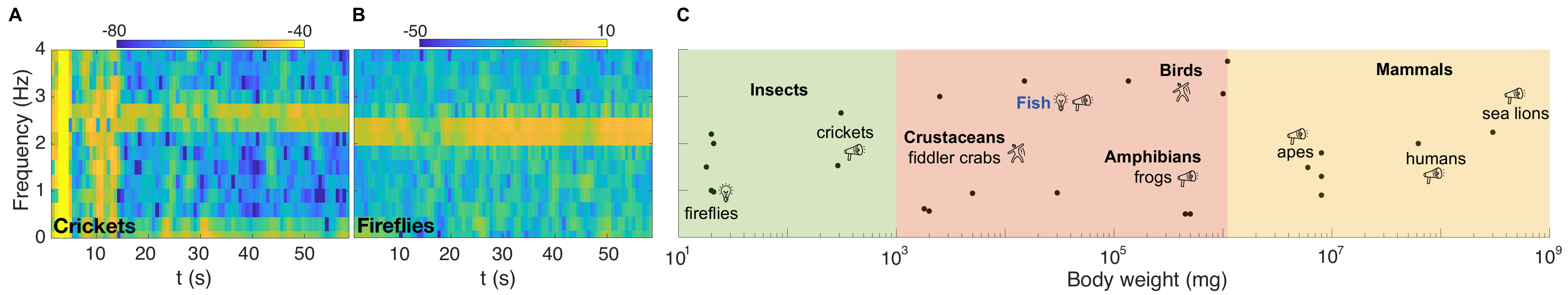}
\caption{\textbf{Tempo comparison across scales, taxa, modalities, and media}. (A) Spectrogram of cricket(s) chirping for 1 min. (B) Spectrogram of nearby fireflies flashing for 1 min ($N=21$). The colorbars in both heatmaps correspond to Power/frequency (dB/Hz). (C) Typical tempos at which different animals signal versus their respective mean body weights on a logarithmic scale. The plot consists of six main groups: insects, amphibians, birds, fish, crustaceans (these last four in an overlapping region due to similar weights---note that the labels here don't correspond necessarily to specific points as the species are mixed), and mammals. The icons (light bulb, speaker, and a moving human) represent the form of the signal (light, sound, or gesture). Note that the signals are mostly transmitted through air, with two examples through water (both fish, written in blue).}
\label{fig:combined}
\end{figure*}

We perform a meta-analysis of peer-reviewed publications covering the animal kingdom (Fig.~\ref{fig:combined}C) and find an abundance of communication tempos in the 0.5-4 Hz range, also known as the ``delta wave'' in neuroscience.  We focus on \textit{isochronous} communication in which signals are repeated multiple times with stable intervals between signal onsets (i.e., metronome-like signals: see Methods for details).

Our data reveal an apparently universal motif. Across 8 orders of magnitude in body weight, across modalities (vision vs audition), and across the tree of life (insects including fireflies\cite{buck1981control,martin2024embracing} and crickets\cite{greenfield2005mechanisms}; crustaceans\cite{backwell1998synchronized,greenfield2005mechanisms}; amphibians\cite{aihara2011complex,greenfield2005mechanisms}; birds\cite{dalziell2013dance,backhouse2024performative,Thomas}; fish\cite{jagers2021social,burchardt2021primer}; and mammals including apes\cite{raimondi2023isochrony,ma2024small}, humans\cite{ito2022spontaneous}, and sea lions\cite{ravignani2017paradox}), communicating at a carrier frequency of $\sim$0.5-4 Hz seems to be the norm. The instances we found are widespread---they come from all around the world and occur in creatures from the air, land, and sea.

We note that there is a risk of selection bias in our data, which could come about in two ways: we may have chosen an unrepresentative set of examples from published literature, and the literature itself could be unreprentative of the full set of biological communication tempos.  It is not possible to fully exclude either of these explanations for our observations, but we did make every effort to avoid inadvertently selecting examples of species consistent with our expectations.  For example, we exclude bird song (e.g., Thrush nightingale and zebra finch) that some consider isochronous but doesn't fit our criteria, though it would also lie within the delta range \cite{roeske2020categorical}. We similarly exclude dog vocalizations around 2 Hz \cite{deaux2024dog}.  However, we do consider bushcrickets, which have been reported to communicate at 11-14 Hz \cite{murphy2016keeping}, and \emph{Saccopteryx bilineata} bat calls at 10-14 Hz \cite{burchardt2019general}, both of which lie outside the delta band (though bats might be expected to communicate differently because of their use of echolocation).

\subsection*{Possible explanation}

The abundance of cases within the 0.5--4 Hz range suggests that there could be some adaptive value to this frequency band. Note that the signalers we've listed are most likely all physically capable of producing signals at higher frequencies (and sometimes do, e.g., \cite{buck1981control}) and obviously could always go slower. Thus they appear to ``opt'' for this frequency band, further suggesting that a communication advantage is present in this range. 

We speculate about the possible underlying reason for this ubiquity. Music psychologists have suggested that humans favor musical tempos in the neighborhood of 2 Hz (or 120 beats per min---BPM) because of their gait \cite{ito2022spontaneous}. Here we argue that this tempo is actually much more widespread, and thus perhaps it is utilized for a different reason. We hypothesize, given the wide variety of signal production apparatuses, that it is likely not a biomechanical constraint on production; the reason could stem from the receiver's end. Most likely this isn't something to do with sensory organs, as again, the channels of communication are diverse. On the other hand, all the receivers rely on some form of neural machinery for processing these signals---this is the common factor. 

Neural circuits across distinct species might share similar characteristics if they had been selected to process similar information from the world. For example, vertebrate visual systems across species show an excess of dark spot detectors (OFF cells) as compared to bright spot detectors (ON cells) because of an asymmetry in the distribution of light in natural scenes \cite{ratliff2010retina}.  Likewise, the rarity of retinal blue cones across species, and the large variance in red to green cone ratio in individual trichromats, can be understood as an adaptation to the structure of chromatic information in natural scenes combined with the constraints of lens-based eyes \cite{garrigan2010design}. In these cases neural circuits behave similarly because they are adapted to the structure of the world.

Another possibility is that neural circuits across species share characteristics because they are adapted to the basic biophysics of neurons.  This may explain, for example, the fact that all mammalian brains show similar neural rhythms \cite{buzsaki2013scaling}.  Indeed, a recent study with rats found that sound patterns in the range of 2 Hz produced the largest neural responses in the auditory cortex \cite{ito2022spontaneous}.

Thus we suggest that the ubiquitous 0.5-4 Hz communication tempo we uncover across scales and taxa may be an adaptation to the resonant frequencies of typical small circuits in the brain. 
Indeed, we show that a sensory ``receiver'' circuit consisting of model neurons with characteristic biophysical integration times of a few hundred milliseconds \cite{izhikevich2007dynamical, dayan2005theoretical, fairhall2001efficiency} will respond best to external stimuli in the frequency range of 0.5-4Hz. 

\subsection*{Computational experiments}

\begin{figure*}[t!]
\centering
\includegraphics[width=8cm]{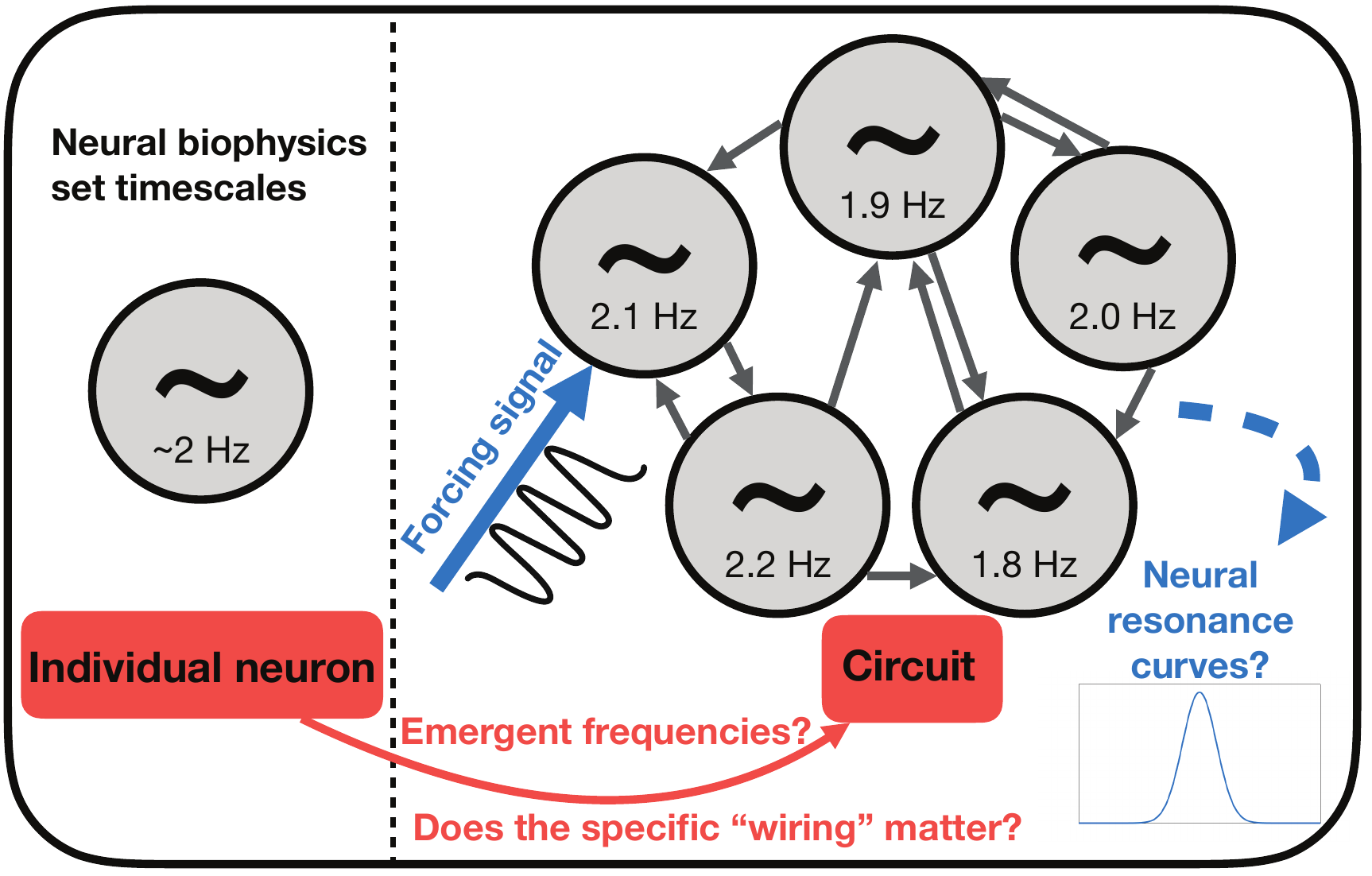}
\caption{\textbf{Schematic of the modeling methodology.} Explanation of the questions we address with our modeling approach: we ask about the ``inheritance" of frequencies from the individual to the circuit level (and if, and how, it depends on the specifics of the connectivity / architecture), and whether we can construct ``neural resonance curves" by forcing our system with an external stimulus (since we measure circuit-level response ($R$), this could also be interpreted as ``circuit resonance").}
\label{fig:schematic}
\end{figure*}

To test our hypothesis that neural circuits in the receiver are best suited for the 0.5-4 Hz frequency range, we perform two computational experiments. First, we investigate whether circuits composed of simple model neurons would typically ``inherit” their fundamental frequency (i.e., will individual neuron integration times set the tempo, or will there will be emergent frequencies, as can be observed in the brain  \cite{buzsaki2004neuronal}?).
Second, assuming that this is indeed the case, we attempt to derive “resonance curves” for these circuits, following the neural-resonance theory framework \cite{large1994resonance, large2009pulse, large2015neural, zalta2024neural}. This demonstrates to what extent these circuits are \emph{entrainable} (i.e., responsive / could be influenced) to external stimuli around 2 Hz \cite{sakaguchi1988cooperative, childs2008stability, antonsen2008external}, and how that depends on parameters.  See  Fig.~\ref{fig:schematic} for a visual explanation of the methodology.

In our experiments, we use an \emph{order parameter}, $R$, to quantifies how ``in sync'' each circuit is with the external signal. This may also be interpreted as a form of ``circuit amplitude,'' since more neurons fire together when the order is higher. See Methods for a detailed explanation of how the order parameter is calculated.

The global coupling (how much the neurons respond to input, whether from another neuron or the forcing) was set so the unforced system was slightly ``subcritical,'' i.e., just below the critical coupling strength for a phase transition\footnote{It is well known in physics that the coupled oscillator models like the Kuramoto model can undergo phase transition from an incoherent (disordered) state to a partially synchronized state when coupling exceeds a critical threshold. That a natural system would evolve to operate near a critical point would be consistent with much work on ``self-organized criticality'' (see, e.g., \cite{bak1988self, de2006self, hesse2014self, rubinov2011neurobiologically}).}. This means that the neurons themselves won’t sync without input---it is explicitly the influence of the forcing that might sync them.

\begin{figure*}[t!]
\centering
\includegraphics[width=\linewidth]{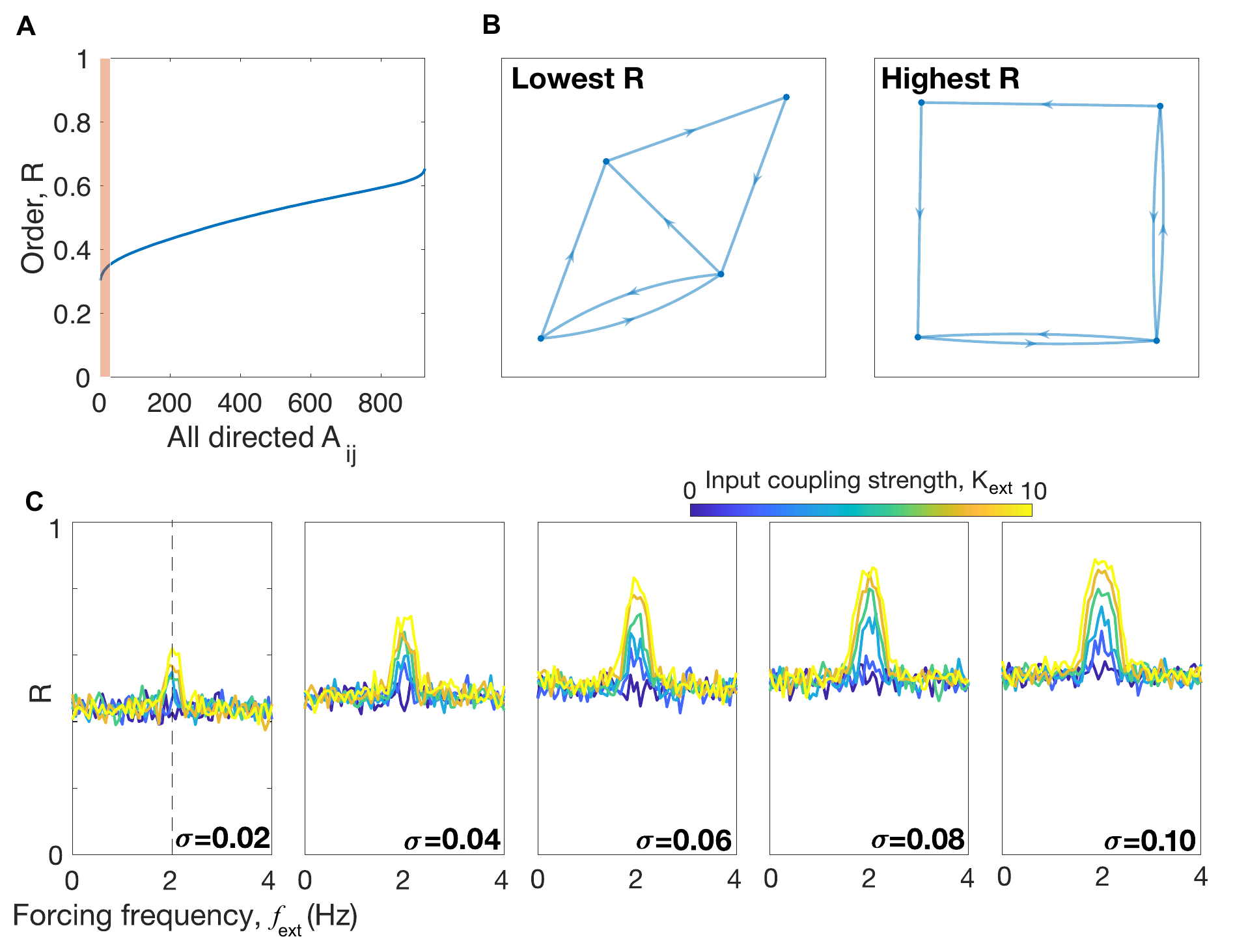}
\caption{\textbf{Modeling circuits of neurons.} Circuits of Kuramoto oscillators with $N=5$ oscillators driven by an external stimulus of 2 Hz. (A) The mean order parameter $R$ (quantifying synchrony) of each circuit (with 10 edges, we ran this for all 1665 possible directed graphs, excluding isomorphic graphs). The data is sorted from the lowest to the highest R. The curve is a mean of 100 curves obtained from 100 realizations. The red shaded area is to draw attention to the lower R regime. (B) Example topologies: we show the topologies of the best and worst in terms of entrainability (highest and lowest R). (C) Neural resonance curves exploring the parameter space of $\Kext$ and $\sigma$ (in this case these are all-to-all topologies). We vary the forcing (i.e., external) frequency $f_{\mathrm{ext}}$ from 0 to 4 Hz. These curves are averages obtained from 10 different realizations (10 different initial seeds). The dashed line in the leftmost plot is at 2 Hz---the mean of the natural frequencies of the circuits.}
\label{fig:models}
\end{figure*}

For the first computational experiment, we exhaustively test (for certain small circuits) all possible topologies to see if the emergent behavior is sensitive to how the circuit is wired. In other words, are some circuits much more or less entrainable?

The assumption here is that neuronal wiring is most likely not fine-tuned.  Fig.~\ref{fig:models}A and B show that, for the vast majority of circuit topologies, response to a 2 Hz driving signal is not sensitive to topology---fine-tuning the graph structure appears to have little effect on the induced order $R$. 

For the second computational experiment, we examine how different forcing frequencies affect the receiving circuit. Fig.~~\ref{fig:models}C shows that a form of resonance is present, and that greater neural diversity allows for stronger induced order even at frequencies different from the mean intrinsic frequency of the neuron population.  This widening of the resonance curve may suggest an evolutionary role for heterogeneity in neurons of the receiver circuit: larger heterogeneity allows for a strong response even to signals that are slightly detuned from the expected external forcing frequency.

\section*{Discussion and Conclusions}

We have demonstrated that many communication signals across the animal kingdom focus on a narrow 0.5-4 Hz frequency band. We propose a mechanistic explanation for this phenomenon. We consider small signal receivers built from neural components with intrinsic timescales set by neural integration times.  We show that (1) these circuits are similarly entrainable regardless of circuit topology, and (2) these circuits will resonate to an external stimulus in a manner akin to resonance in physical systems, though, unlike physical systems, they resonate more robustly with a more diverse set of component neurons.

We intend this work as a starting point for further research. As such, we have not yet explored all implications of our model choices and possible alternatives and generalizations. Some key choices in our model include the type of oscillator (Kuramoto model without inhibition), the circuit size ($N=5$; though we do show, as a proof of concept, that these results can scale), and the coupling matrix structure (binary nonsparse). We numerically explore alternative choices to test that our conclusions appear robust, but leave thorough examination of the many options for future work.

Indeed, it would be interesting to leverage the significant corpus of work on different models of coupled neurons to explore the robustness with which physical properties of neurons set timescales for resonant synchronous excitation of circuits.

We also note that the frequency band of 0.5-4 Hz corresponds to the delta wave, the lowest frequency brain rhythm commonly identified \cite{buzsaki2006rhythms}. In physical systems, harmonic resonance---where a lower forcing frequency excites a higher frequency response---is much more easily achieved than subharmonic resonance, where the reverse occurs \cite{likharev2013essential}. By analogy, one might expect that an external stimulus exciting the lowest frequency band of brain waves would allow it to more easily couple to and influence all of the higher frequency rhythms in the brain.

More empirical observations (as also suggested by Hersh et al. \cite{hersh2023robust}) could help shed light on how omnipresent this tempo indeed is, or whether there may even be other tempo ``hotspots.''  For instance, we included gesturing as a form of communication. The generation of such signals will likely change with body size---when limbs become larger, they move more slowly; compare a cockroach's run with an elephant's gait \cite{ord2023evolutionary}. We included  gestures from fiddler crabs and a limited number of bird species (which are, roughly speaking, similar body sizes). A possible implication of our work is that isochronous gesturing should not be prevalent in animals that are much larger or much smaller as they would fall outside the 0.5-4 Hz band for mechanical reasons.

\section*{Methods}

\subsection*{Field data}

Cricket and firefly data presented in Fig.~\ref{fig:combined} come from analysis of original video and audio recordings collected with a Sony Alpha 7 SII camera in Amphawa, Thailand in summer 2022. 

Firefly flash timings were obtained via a computer vision algorithm we developed as described in \cite{PopulationMonitoring}. To obtain a global picture of synchronous firefly frequencies, we summed all flashes per frame (we want the overall group dynamics) and then smoothed this signal (Savitzky-Golay filter of polynomial order 2 and frame length 9), subtracted the mean, and plotted the spectrogram (time window 100 samples = 3.33 seconds for 29.97 FPS video). All computational scripts described here and throughout were written in Matlab v2020a \cite{MATLAB}.

For the cricket (audio) data we first applied a highpass filter (cutoff 5 kHz) then computed the peak envelope of the absolute value of the filtered data using a Hilbert filter of window size 3000 samples (68 ms). The mean was then subtracted and spectrogram plotted.

\subsection*{Data from previously published work}

We began our meta-analysis with a search for widely accepted isochronous animal signals.  However, because the definition of isochrony varies in the published literature, challenges immediately arose.  For example, for inclusion in our dataset, it was necessary to establish a minimum number of signal repetitions (5) and a minimum level of IOI (inter-onset-interval) consistency (25\%).  Neither of these are typically made explicit in relevant publications, so there was often a need for manual estimation from figures or close reading of text and tables. Furthermore, some isochronous (or quasi-isochronous) signals are not flagged as such in relevant publications.  Therefore, our dataset is very likely non-exhaustive, and automating the processes seems out of reach at this point.

We wished to test our hypothesis of the universality of the 0.5-4Hz communication band by seeing if and when it broke down: are there limitations related to body sizes, taxa, or communication modalities?  This motivated collection of data across the broadest possible range of species.

In Fig.~\ref{fig:combined} we show approximate values for the weights and tempos of the given species. Exact weights and tempos were rarely reported in the referenced works. For weights, we estimated based on searches for typical values for each species, ignoring distinctions between males and females.  We note that, even if our estimated weights were off by a factor of two, our figure would remain largely unchanged due to the many orders of magnitude covered. For tempos, we approximated frequencies from figures and/or data shared in each study (e.g., a time series plot).

\subsection*{Computational experiments}

To test our hypotheses we simulate the Kuramoto model \cite{kuramoto1975self, strogatz2000kuramoto, strogatz2024nonlinear}. It is well known that neurons respond to synaptic currents by spiking intermittently and often rhythmically; this spiking can be regarded a form of oscillation and modeled at different levels of accuracy via systems of differential equations. The Kuramoto model offers simplicity and yet it was used previously to model neural circuits---see, e.g., \cite{breakspear2010generative, cumin2007generalising}. It can be seen as a limit of the Stuart-Landau model---recently utilized by Zalta et al. \cite{zalta2024neural}---where amplitudes are nearly constant.

We modeled circuits of Kuramoto oscillators according to

\begin{equation}
 \dfrac {d\theta_i} {dt} = \omega_i + \frac{K}{N} \left[\sum_{j=1}^{N} A_{ij} \sin({\theta_j - \theta_i}) + \Kext \sin(\thetaext-\theta_i)\right], \indent{i=1,...,N} .
 \label{eq:forcedKM}
\end{equation}
where adjacency matrix $\mathbf{A}=[A_{ij}]$ defines the coupling circuit, $\theta_i$ and $\omega_i$ represent the internal phase and natural frequency of the $i$th oscillator, respectively, $N$ is the number of oscillators, $K$ is the inter-neuron coupling strength, and $\Kext$ is the strength of coupling to the external forcing (expressed in units of $K$ so that $\Kext=1$ corresponds to external forcing being equal in magnitude to internal coupling).

The critical coupling $K_c$ was determined using the large-$N$ formula $K_c = 2  [\pi g(0)]^{-1}$ (exact as $N \to \infty$), where $g(\omega)$ is the probability distribution function (PDF) for the oscillator natural frequencies.  When the natural frequencies are normally distributed with standard deviation $\sigma$, this becomes $K_c = 2\sigma\sqrt{2/\pi}$.  In practice we fixed $K$ just below this theoretical critical point at $K = (1-\epsilon) K_c$ with $\epsilon=0.1$. Note that the fact that $K_c$ depends on $\sigma$ means that the critical coupling changes as the oscillator diversity changes.

Numerical integration of system \eqref{eq:forcedKM} was performed with an explicit fourth order Runge-Kutta method (ode45 in Matlab). We ran multiple realizations where, in each realization, both oscillator initial phases and natural frequencies were selected randomly, with phases chosen uniformly at random and frequencies chosen from a Gaussian distribution of mean $\mu$ and standard deviation $\sigma$. When we varied the graphs $A_{ij}$ we kept the mean frequency fixed at $\mu=2$ Hz.

We use the order parameter $R$ to quantify (on a zero to one scale) the synchrony or ``orderliness'' of the group at each point in time \cite{strogatz2024nonlinear}:
\begin{equation} \label{eq:basicR}
    R(t) = \left| \left< e^{i \theta_j} \right> \right| = \left| \frac{1}{N}  \sum_{j=1}^N e^{i \theta_j}  \right|.
\end{equation}
This can be understood as the distance to the center of mass of all oscillators if one imagines each one represented by a point-like mass located on the unit circle at an angle corresponding to its phase $\theta_i$. Then the ``no sync'' state (where oscillators have random or uniformly distributed phases around the circle) will be located at the origin of the coordinate system, while the center of mass for the ``perfect sync'' state (where all oscillators have the same phase) will lie on the unit circle.

Note that we calculate $R$ including the phase of the external forcing (i.e., treating it as a sixth node); by doing so we incorporate information about entrainment to the forcing and not just synchronization of the neurons (higher $R$ means the forcing and the circuit are in sync). 

In this study, we choose to focus on circuits of five vertices and exactly ten directed edges (a total of 1665 unique non-isomorphic circuits). This is because the number of graphs grows super-exponentially, and thus full computational exploration of all graphs of size $N$ quickly becomes prohibitively expensive. We exclude self-coupling (loops of length 1) and choose a fixed value of ten edges to maximize the number of distinct circuits covered and focus on the coupling structure rather than the dependence of entrainment/resonance on number of edges.

When deriving ``neural resonance curves,'' we fix the coupling topology to be global (all-to-all, i.e., fully connected) and evaluate the effects of varying both the heterogeneity of the oscillators, quantified by standard deviation $\sigma$, and the magnitude (relative weighting) of the forcing $\Kext$.  As Fig.~\ref{fig:models}C shows, we found that the response peaked at the mean oscillator frequency and that increasing $\Kext$ and $\sigma$ both widened and amplified the peak.

\section*{Acknowledgments}
We thank Michael Greenfield, Thomas Stoeger, Itamar Lev, Thomas MacGillavry, Hermann Riecke and Bard Ermentrout, for insightful discussions that greatly improved the paper. We thank Anton de Lesseps for audio data preparation. We also thank the entire 2022 field team for their support, especially Robin Meier Wiratunga, Emma Zajdela and Tanthai Prasertkul. G.A. and D.M.A. acknowledge support from the Buffett Institute for Global Affairs (Northwestern University), Northwestern Institute on Complex Systems (NICO) as well as the National Institute for Theory and Mathematics in Biology (NITMB). V.B. is supported in part by the Eastman Professorship at Balliol College, University of Oxford.


\bibliography{2HzArXiv}

\newpage
\clearpage
\renewcommand\figurename{\textbf{Supplementary Fig.}}
\setcounter{figure}{0}
\renewcommand{\thefigure}{S\arabic{figure}}

\section*{Supporting Information: ``A universal animal communication tempo resonates with the receiver's brain''}

\subsection*{Extended neural resonance results} 
\begin{figure}[h!]
\centering
\includegraphics[width=0.5\textwidth]{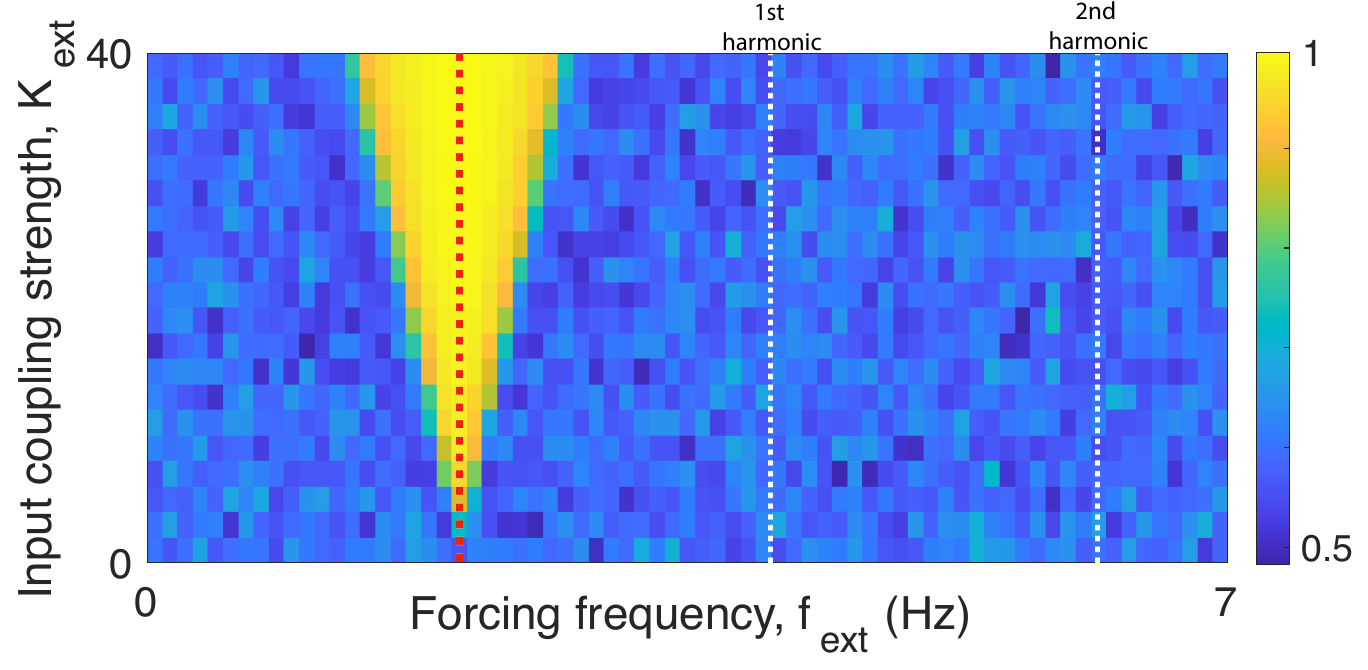} 
\caption{\textbf{Arnold tongues.} A heatmap showing the response of the system as a function of the external driving amplitude and the forcing frequency. This reveals a ``tongue" at around 2 Hz, that widens as the input strength increases. We see no additional tongues at the harmonics (highlighted with white dashed lines) or subharmonics. Result obtained from 10 realizations (initial seeds).}.
\label{fig:ArnoldTongue}
\end{figure}

\begin{figure}[h!]
\centering
\includegraphics[width=0.5\textwidth]{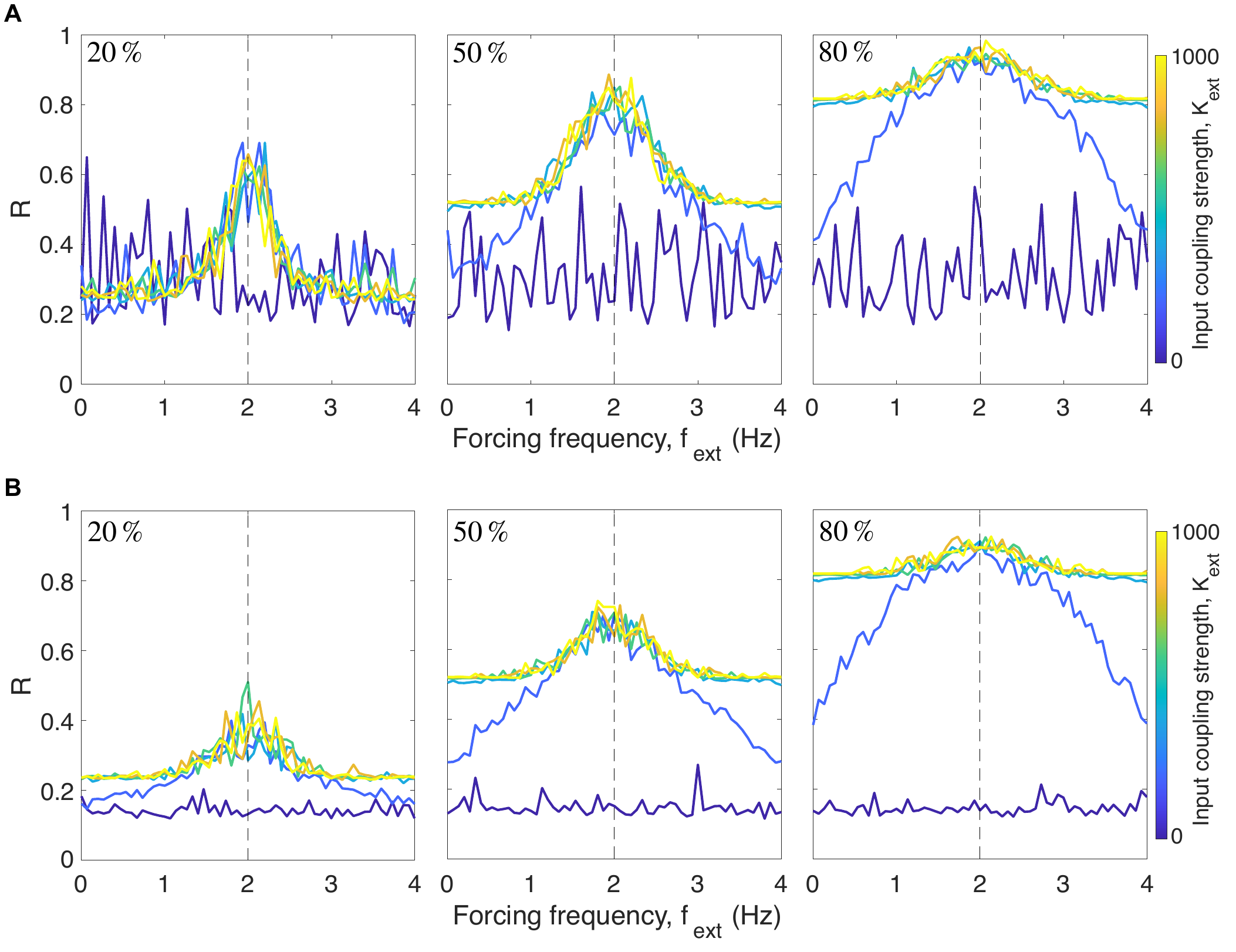}
\caption{\textbf{Resonance with a large network.} Resonance curves with $N=100$. We vary the fraction of input neurons (with 20\%, 50\% and 80\% in separate plots from left to right). Results obtained from 1 realization (initial seed). (A) All-to-all network. (B) A random binary network, with each node assigned 60 random connections to avoid sparsity.}.
\label{fig:InputFractions}

\end{figure}

We plot Arnold tongues for this system (Fig.~\ref{fig:ArnoldTongue}); we observe only one tongue at 2 Hz, with no additional tongues at harmonics or subharmonics (as would be expected, e.g., for mechanically resonant systems). Finally, to check the robustness of our results in larger networks, we also perform simulations with $N = 100$ with varying fractions of input neurons (SI Fig.~\ref{fig:InputFractions}); we observe similar resonance curves both with all-to-all and random binary networks.

\subsection*{Extended graph topology results} 
When we attempt to ask what sets the ``weaker'' graph topologies apart from the others, we find that there are more reciprocal pairwise connections (i.e., if A is connected to B, B is also connected to A). In this 5-neuron example, the 10 lowest order cases have a mean of 5.38 reciprocal edges (i.e., graph reciprocity 0.54) as opposed to 5.08 in the 10 highest order cases (i.e., graph reciprocity 0.51): see Fig. \ref{fig:models}B of the main text for example graphs. The graphs also differed in degree of clustering, with a higher clustering coefficient in the lower order graphs (0.38 vs 0.32).

\end{document}